\documentclass[runningheads]{llncs}
\usepackage[T1]{fontenc}
\usepackage{graphicx}
\usepackage{multicol}
\usepackage{enumitem} 
\providecommand{\tightlist}{%
  \setlength{\itemsep}{0pt}\setlength{\parskip}{0pt}}

\usepackage{orcidlink}

\begin{document}
\title{Security in the Era of Perceptive Networks: A Comprehensive Taxonomic Framework for Integrated Sensing and Communication Security}
\titlerunning{Security in the Era of Perceptive Networks}
%
%
\author{Chandra Thapa\orcidlink{0000-0002-3855-3378} \and
Surya Nepal\orcidlink{0000-0002-3289-6599}}
\authorrunning{Thapa et al.}
%
\institute{CSIRO Data61, Sydney, Australia \\
\email{chandra.thapa@data61.csiro.au; surya.nepal@data61.csiro.au}\\
}
\maketitle              
%

\begin{abstract}
Integrated Sensing and Communication (ISAC) represents a significant shift in the 6G landscape, where wireless networks both sense the environment and communicate. While prior comprehensive surveys have established foundational elements of ISAC security, discussed perception-focused security models, and proposed layered defense strategies, this paper synthesizes these studies into a comprehensive taxonomic framework that covers the whole ISAC security domain. 
This paper provides a systematic and thorough review of ISAC security across multiple orthogonal dimensions. These include threat taxonomy and propagation methods; vulnerability analysis at design, physical, computational, and architectural levels; defense mechanisms categorized by deployment layer; security-performance trade-offs with theoretical bounds; sector-specific security demands for critical infrastructure; and emerging issues such as quantum resilience, AI-hardening, and privacy preservation. Unlike previous frameworks that primarily focus on vision, this review combines these dimensions, introduces new classification schemes that reveal hidden relationships between threats and defenses, and identifies key research gaps through structured analysis. This detailed taxonomy offers a valuable reference for researchers developing secure ISAC systems and policymakers establishing security standards.
\end{abstract}
%
\section{Introduction}
The trajectory of wireless communication systems has historically been characterized by a singular pursuit: optimizing data transmission. However, as the industry progresses toward the Sixth Generation (6G), a fundamental paradigm shift is taking place: the network is no longer merely a conduit for information but also functions as an intelligent sensor of the physical environment~\cite{aman2025}\cite{huawei_wp}. This evolution gives rise to perception-centric networks that actively observe and interpret the physical surroundings in real time to inform immediate control decisions in critical systems, such as transportation and healthcare~\cite {huawei_wp}.

At the heart of this vision is Integrated Sensing and Communication (ISAC), which integrates radar sensing and wireless communication into a unified infrastructure that shares hardware, spectrum, and processing resources~\cite{matsumine2025}. 
Its importance is high in 6G as the International Telecommunication Union (ITU) recognizes ISAC as one of six core usage for 6G (IMT-2030)~\cite{ITU_reco,5Gamerica_whitepaper}. ISAC systems can integrate sensing and communication at three increasing levels of sophistication~\cite{liu2022}:
(1) \textbf{Spectrum sharing (coexistence-level integration):} Radar and communication operate in the same frequency band, but with distinct hardware and waveforms; (2) \textbf{Hardware sharing:} A single platform transmits and receives for both functions, potentially time-multiplexing tasks switching between radar mode and communication mode; and (3) \textbf{Waveform sharing (full integration):} A unified waveform simultaneously carries information and probes the environment (primary focus for 6G). 
    
    

By unifying these functionalities on shared hardware and spectrum platforms, ISAC promises significant efficiency gains and cost reduction~\cite{ISAC3,huawei_wp}. More importantly, ISAC enables novel capabilities, such as cooperative sensing for autonomous vehicles~\cite{huawei_wp}, enhances the communication infrastructure~\cite{huawei_wp}, distributed environmental monitoring for smart cities, and real-time physical security for critical infrastructure~\cite{smart_grid1}, that require seamless integration rather than separate systems~\cite{thapa2025}.

The dual-function nature of ISAC, where a single waveform carries both communication and sensing information, introduces novel security risks because the sensing targets themselves can act as potential eavesdroppers~\cite{PLS_AN}.
The broadcast nature of wireless signals is at the core of this problem, as it inherently exposes confidential data embedded within the sensing waveform to these targets~\cite{matsumine2025,PLS_AN,Guo2024}.
Moreover, the transformative capabilities that enable, for example, the detection of unauthorized drones can also be exploited for warrantless surveillance~\cite{thapa2025}. The adversarial objective fundamentally shifts from data theft to manipulating the network's perception of reality. This way, with ISAC, a new security paradigm centered on perception integrity is required to safeguard the network's understanding of reality. To capture the essence of the security evolution, Table~\ref{tab:tvsc} provides a direct comparison between the traditional cybersecurity paradigm and the emerging cyber-physical security paradigm needed for ISAC.

While existing literature addresses ISAC security components in isolation, such as physical-layer security (PLS) and Reconfigurable Intelligent Surfaces (RIS)-assisted defenses, a comprehensive taxonomic framework that unifies the ISAC security landscape remains needed. A systematic and holistic review is required to characterize the whole security environment, including threat taxonomy, vulnerability root causes, defense mechanisms, performance trade-offs, requirements, and emerging research gaps.

\begin{table*} [ht]
    \centering
    \caption{Traditional vs. ISAC-centric security paradigms.}
    \scriptsize
    \label{tab:tvsc}
    \begin{tabular}{|p{1.5cm}|p{4.8cm}|p{5.5cm}|}
    \hline
    \multicolumn{1}{|c|}{\textbf{Attributes}} &
    \multicolumn{1}{|c|}{\textbf{Traditional cybersecurity} }&
    \multicolumn{1}{|c|}{\textbf{ISAC cyber-physical security} }
        \\ \hline
         Primary target & Digital data: Confidentiality, integrity, availability & Physical system perception: accuracy and trustworthiness
         \\ \hline
         Attack domain & Digital intruder: Network, software layers, cryptographic systems  & Cyber-physical: RF spectrum, signal processing, control loops 
         \\ \hline
         Key attack vectors & Malware, phishing, network intrusion, DDoS, man-in-the-middle  & Radio frequency (RF) spoofing, deceptive jamming, CSI manipulation, signal injection 
         \\ \hline
         Adversary's goal & Steal, encrypt, or corrupt digital data  & Deceive control systems, induce physical consequences 
         \\ \hline
         Detection basis & Digital signatures, logs, cryptographic verification & Physical signal anomalies, environmental consistency checks
         \\ \hline
         Encryption & Validates data authenticity in transit/at rest & Cannot validate the integrity of physically manipulated signals; it may be too latent for real-time control loops 
         \\ \hline
         Trust model & Perimeter-based (trusted inside, untrusted outside) & Zero-trust (never trust, always verify, continuous authentication)
         \\ \hline
    \end{tabular} 
    \vspace{-0.2cm}
\end{table*}

\subsection{Contributions}
This paper provides a foundational reference by moving beyond conceptual visions to establish a formal, multi-dimensional taxonomy for ISAC security, synthesizing the field across all necessary dimensions. The contributions are:
\begin{enumerate}[leftmargin=*]
    \item \textbf{Comprehensive threat taxonomy:} Classification of all known ISAC threats into multiple fundamental categories, introduction of threat propagation taxonomy, and characterization of multiple specific attacks with threat models.
    \item \textbf{Vulnerability hierarchy and root cause analysis:} Classification of vulnerabilities into four levels, mapping them to fundamental constraints, and identification of unavoidable versus preventable root causes.
    \item \textbf{Defense mechanism taxonomy:} Classification of defense mechanisms into four synergistic pillars, provision of cost-benefit analysis, and identification of complementary versus redundant defenses.
    \item \textbf{Security-performance trade-offs:} Quantitative analysis of formal bounds relating communication rate, sensing accuracy, and information leakage.    
    \item \textbf{Sector-specific security requirements:} Analysis of six critical infrastructure sectors: smart grids, intelligent transportation, healthcare, Unmanned Aerial Vehicle (UAV), industrial systems, and smart cities, including threat models and required defenses for each.
    \item \textbf{Research gaps and emerging challenges:} Analysis of research clusters and gaps, including defense mechanisms, and prediction of emerging challenges (quantum, advanced Artificial Intelligence, privacy).
\end{enumerate}

\section{Comprehensive threat taxonomy}
The ISAC threat landscape is multidimensional, extending far beyond simple jamming to include sophisticated manipulation of the cyber-physical interface. We categorize threats based on their operational mechanics, their targets, and, crucially, their propagation dynamics.
\subsection{Taxonomy by attack types}
The first dimension categorizes attacks based on their intended goal and operational style. Attack type is divided into two dimensions:

\subsubsection{Role-based classification}
This classification categorizes the role of wireless signals in the security scenario~\cite{geng2024}.
\begin{itemize}[leftmargin=*]
    \item \textbf{(Category 1) Signals as victims:} The focus is on disrupting target estimation by manipulating signal sources (jamming, spoofing) or channels. Attacks on wireless sensing systems (e.g., automotive radar or human activity recognition systems) render them inoperable or inaccurate.
    \item \textbf{(Category 2) Signals as weapons:} Wireless signals are used to attack other systems, stealing information or causing malfunction, often for malicious information estimation (e.g., measuring physiological states). This includes active attacks (in which the attacker transmits probe signals) and passive attacks (in which the attacker listens to existing signals).
    \item \textbf{(Category 3) Signals as shields/guardians:} Wireless signals are actively used for security applications, such as physical layer authentication or data protection.
\end{itemize}

\subsubsection{Operational classification}
\begin{itemize}[leftmargin=*]
    \item \textbf{Passive attacks}: Passive attacks involve gathering intelligence and monitoring communications without directly interfering with the signal integrity~\cite{thapa2025}. Examples include eavesdropping on communication data or sensing information, and performing traffic analysis~\cite{Guo2024,keskin2025}. Passive sensing eavesdroppers exploit the transmitted ISAC waveform reflected by targets to extract sensing information regarding those targets or their surrounding environment, posing security and privacy challenges~\cite{matsumine2025,PLS_AN}.
            
    \item \textbf{Active attacks} Active attacks involve deception and disruption~\cite{thapa2025}. These attacks include jamming, which is the intentional transmission of interfering signals to reduce the signal-to-noise ratio (SNR) at legitimate receivers~\cite{RISbasedPLS,cigno2025}. Spoofing involves generating false signals, commands, or data that deceive the target system into interpreting them as genuine~\cite{target_spoofing,geng2024,Guo2024}. Active attacks can also involve the deliberate manipulation of Channel State Information (CSI) by injecting false data or pilot contamination~\cite{keskin2025}, or manipulating the wireless environment through technologies such as RIS by applying analog phase shifts to rapidly alter signal reflections~\cite{Naeem2023,Xu}.
\end{itemize}

\subsection{Taxonomy by attack target}
The second dimension focuses on which specific component of the ISAC system is compromised.    
    \begin{itemize}[leftmargin=*]
        \item \textbf{Communication and sensing channels:} Attacks target both the communication channel, aiming to compromise data confidentiality, and the sensing channel, aiming to compromise the integrity of sensing information~\cite{PLS_AN,geng2024}. The tight coupling inherent in ISAC means that a security compromise of one channel can propagate to the other~\cite{keskin2025}.
        
        \item \textbf{Artificial Intelligence/Machine Learning (AI/ML) models:} The reliance on AI/ML for functions like beam prediction and data fusion makes the models vulnerable to adversarial attacks, such as data poisoning during training or evasion attacks during inference~\cite{keskin2025,Nguyen,son2025}.
        
        \item \textbf{Cross-layer dependencies and system coordination:} Compromising components like sensing units can introduce corrupted data into real-time control loops, leading to inappropriate protocol-level decisions in subsequent network layers~\cite{keskin2025,surve2025}. This includes the compromise of infrastructure-based ISAC sensing, which feeds directly into onboard perception and control algorithms~\cite{keskin2025}.
    \end{itemize}
    
\subsection{Taxonomy by attack scope}  
This dimension describes the geographic or systemic breadth of the attack:
    \begin{itemize}[leftmargin=*]
        \item \textbf{Single-node attacks:} These attacks focus on exploiting individual devices or base stations, such as an attacker sniffing data transmitted by a single legitimate user.

        \item \textbf{Network-wide propagation, cross-infrastructure cascades, and cascading failures:} Due to the complex coupling and integrated nature of ISAC networks (e.g., vehicular networks), a fault originating in one node can propagate across the network~\cite{seccure_isac1,surveysecuritysmart}. The complexity of humanoid systems, for example, means that a local compromise can cascade through sensing, planning, and actuation layers. Such cascading failures are a key security concern~\cite{keskin2025}.
    \end{itemize}

\subsection{Taxonomy by threat propagation}
Threats in ISAC propagate systemically due to the inherent tight coupling of sensing and communication functionalities, which creates critical dependencies that allow a security compromise in one area to cascade into others. This propagation can be categorized into four types:

\textbf{(1) Vertical propagation (cross-layer):} Vertical propagation describes the rapid failure mode where an attack originating in a lower architectural layer, such as the physical layer, compromises a higher-level computational system, leading to a physical consequence~\cite{surve2025}. In ISAC systems, tighter cross-layer coupling leads to faster error propagation across layers~\cite{keskin2025}. A physical-layer attack, such as Channel State Information (CSI) manipulation~\cite{CSI_1}, can corrupt the raw sensing data used in computational processes, causing errors or confusion in Artificial Intelligence (AI) detectors~\cite{keskin2025}. This results in corrupted input data being fed into real-time control loops, ultimately leading to cyber-physical domain failures, such as unsafe actuation or inappropriate protocol-level decisions~\cite{keskin2025}. Because robotic and autonomous control loops operate under strict time constraints, the total latency from the initial physical attack to the resulting safety consequence can be as low as a few seconds~\cite{surve2025}. 
  
\textbf{(2) Horizontal propagation (network-wide):} Horizontal propagation occurs when an attack spreads laterally across a network by leveraging shared infrastructure, compromising multiple independent nodes~\cite{Guo2024}. This is especially prevalent in cooperative environments, such as Vehicle-to-Everything (V2X) networks, where shared sensor data and communication links are essential for coordination~\cite{cui2025,ma2024location}. For instance, a spoofing attack targeting a single vehicle in a V2X platoon can introduce erroneous position information, which is then communicated via the V2X link to other trucks in the platoon~\cite{cui2025}. Acting on this corrupted data, the system may trigger platoon-wide emergency braking, potentially causing a catastrophic pileup~\cite{ma2024location}.
    
\textbf{(3) Cascading propagation (system interdependencies):} Cascading propagation involves an attack on one critical infrastructure system leading to consequential failures in logically or physically dependent systems~\cite{Guo2024}. Smart grids are highly susceptible to this, in which compromising a component can lead to complex failures that affect both the cyber and physical domains~\cite{Huang_2025}. For example, grid voltage manipulation or command disruption in control systems could trigger wide-area instability or physical damage, thereby affecting dependent communication or Internet of Things (IoT) devices that rely on the grid's operational integrity~\cite{surveysecuritysmart,Huang_2025}. The complexity of these attacks makes mitigating them difficult, as they bypass defenses focused solely on a single layer~\cite{surveysecuritysmart}.
    
\textbf{(4) Temporal propagation (time-delayed impact):} Temporal propagation involves attacks designed to remain latent within a system over an extended period before activation~\cite{Nguyen}. The most common mechanism for this is data or model poisoning in AI training pipelines, particularly in distributed learning environments such as Federated Learning (FL)~\cite{son2025}. The attack injects malicious data or corrupted model updates during training~\cite{Nguyen}. These modifications remain hidden within the model until environmental or input conditions align, at which point the malicious logic, often a backdoor, is activated, manifesting the attack's effects over a latency ranging from hours to months~\cite{surve2025}.

\section{Vulnerability hierarchy and root-cause analysis}

\subsection{Vulnerabilities classification}
Understanding vulnerabilities helps design effective defenses for the ISAC system. These vulnerabilities are classified into four hierarchical levels, from design choices to processing issues.

\textbf{(Level 1) Design-level vulnerabilities (fundamental choices):} These vulnerabilities stem from the fundamental choices in system design. The selection of standardized, predictable waveforms, such as OFDM, in communication\allowbreak-centric designs inherently renders systems highly vulnerable to advanced spoofing and deceptive jamming attacks. Standardized WLAN signals and protocols are exploited for radar-like sensing, making the system sensitive to deceptive jamming~\cite{deceptive_jamming,target_spoofing}. Since OFDM allows the generation of digitally synthesized range/Doppler maps, it is vulnerable to deceptive jamming~\cite{deceptive_jamming}. The dual functionality of ISAC signals, particularly standardized pilots in Wi-Fi sensing, makes them susceptible to spoofing attacks~\cite{keskin2025}. 

\textbf{(Level 2) Architectural vulnerabilities (system design):} System architectures introduce trade-offs that create vulnerabilities. A Monostatic Architecture (co-located Tx/Rx) experiences considerable self-interference (SI) that adversaries can exploit via precise timing attacks to inject false targets undetected. The monostatic architecture requires a co-located transmitter and receiver~\cite{cui2025,ISAC_type2}, leading to unavoidable self-interference~\cite{ma2024location,ISAC_type2}. This issue is so significant that strong residual SI in Monostatic ISAC can dominate the radar echo, making high-quality implementation difficult~\cite{bazzi2024}. 
    The architectural reliance on programmable intelligent surfaces (like RIS) introduces the risk of an adversary injecting a compromised device or manipulating a trusted one to perform unauthorized control. This vulnerability, termed Malicious/Illegal RIS (IRIS) Injection, enables passive signal reflection for signal leakage (eavesdropping) or active signal modulation for interference attacks~\cite{Naeem2023}. If the network lacks robust authentication for RIS control, an adversary can manipulate the reflection coefficients, thereby degrading CSI and sabotaging link reliability~\cite{RISbasedPLS}.
    
\textbf{(Level 3) Physical-layer vulnerabilities (RF domain):} Vulnerabilities intrinsic to the RF domain include issues with beam steering and channel estimation. Analog phase-shift beamforming introduces Side-Lobe Leakage, creating unintended eavesdropping zones where adversaries can intercept communication without precise beam alignment. Beam squint effects cause spatial leakage, in which energy intended for a user is misdirected toward nearby angles, thereby expanding the eavesdropping zone, especially in wide-band OFDM systems. Low sidelobe levels are necessary for ISAC to prevent the illicit interception of communication information embedded in signals~\cite{bazzi2024,aman2025}. If these sidelobes are not adequately managed, they can create unintended eavesdropping zones through leakage~\cite{bazzi2024,aman2025}. Beam squint occurs because analog components in wideband systems use the same weights across all subcarriers, causing beams to illuminate distinct, unintended directions, thereby degrading communication capacity and sensing accuracy~\cite{elbir2024}.
    
\textbf{(Level 4) Computational vulnerabilities (processing):} This level concerns the processing of signals. In-sensor computing (ISC) uses analog or mixed-signal circuits, creating an analog gap in which data is processed before digitization. Attackers can inject analog signals that bypass digital validation or use side-channel attacks (electromagnetic emissions) to leak information. In-sensor computing systems integrate sensing and analog computation units, processing data before digitization~\cite{kajol2025}. This integration creates vulnerabilities that allow attackers to inject faults, for instance, by inserting low-cost circuits into the serial data bus to corrupt images before they are processed~\cite{kajol2025}. Because ISC introduces additional analog signals, existing digital countermeasures may be ineffective against new analog faults disguised as noise. These systems are also susceptible to side-channel attacks, including electromagnetic emissions, which can leak sensitive information~\cite{geng2024}.

\vspace{-0.5cm}

\subsection{Root-cause analysis}
Root-cause analysis has two types: unavoidable and preventable.

\textbf{Unavoidable vulnerabilities}
Unavoidable vulnerabilities are rooted in fundamental limitations:

\textbf{(1) Information-theoretic limits:} 
    Eavesdropping on sensing waveforms is fundamental, as ISAC signals are exposed by the broadcast nature of wireless channels, allowing adversaries to directly extract geometric information such as range, Doppler, and angle from the received waveform~\cite{keskin2025,matsumine2025,Guo2024,cigno2025}. Jamming resistance is bounded by the power budget ratio because optimization problems related to resisting attacks are constrained by transmit power limits, and jamming success is defined by power ratios~\cite{Guo2024,keskin2025}. Channel State Information (CSI) estimation accuracy is fundamentally limited by channel coherence time~\cite{liu2022}.

\textbf{(2) Hardware constraints:} 
    Beam squint in mmWave is fundamental to antenna array physics, occurring because hybrid beamforming architectures utilize subcarrier-independent analog components whose weights are set according to the carrier frequency~\cite{elbir2024}. 
    While this root cause is inherent, its deleterious effects can be managed by hardware solutions such as True-Time Delayers (TTDs) or signal-processing corrections~\cite{elbir2024}.
    Analog front-end vulnerabilities are inherent to mixed-signal design, as In-Sensor Computing (ISC) processes data before digitization using analog or mixed-signal circuits, making existing digital countermeasures ineffective against new analog faults disguised as noise~\cite{kajol2025}. Hardware switching-speed constraints limit the reconfiguration latency of Reconfigurable Intelligent Surface (RIS) systems, as RIS hardware is subject to physical constraints on switching rate, resolution, and phase noise~\cite{hafeez2025}.

\textbf{Preventable Vulnerabilities}
Preventable vulnerabilities can be mitigated through careful design choices:

\textbf{(1) Design choices:} Waveform predictability can be addressed through randomization, as randomly distorting signals makes the channel appear ever-changing and invalidates localization techniques based on CSI fingerprinting~\cite{cigno2025,CSI_1,Guo2024}. CSI pilot patterns can be securely handled by using pseudo-random sequences whose seeds are exchanged securely, and beamforming side-lobe energy can be minimized with precoding to prevent unauthorized interception of confidential information embedded in signals~\cite{cigno2025,geng2024,aman2025,xu2022}. Furthermore, the lack of cryptographic integrity verification for RIS control signals is a design choice that exposes the network to Malicious RIS Injection~\cite{RISbasedPLS}.

\textbf{(2) Architectural decision:} Implementing a distributed architecture can eliminate single points of failure, improving resilience, as seen in systems like Cell-Free Massive MIMO~\cite{Guo2024,Nguyen}. Centralized processing can be decentralized through approaches such as edge learning, distributing tasks across multiple edge devices, and shifting from single-point coverage to ubiquitous sensing networks~\cite{cui2025,ISAC_types}. A lack of authentication can be remedied by adding cryptographic mechanisms, such as developing a dual-ID mechanism (digital and physical) to thwart location-sensitive spoofing attacks and using physical-layer and cryptographic checks for session authentication~\cite{Guo2024,keskin2025}.

\section{Defense mechanism taxonomy}
ISAC requires a cross-layer defense-in-depth framework built on four synergistic pillars, emphasizing layered redundancy, and cross-layer orchestration and synergy~\cite{thapa2025}.

\subsection{Four pillars}
\subsubsection{Pillar 1}
\textbf{Physical-layer-aware security} leverages wireless channel properties for secure transmission without relying only on higher-layer encryption~\cite{RISbasedPLS,matsumine2025}.
\begin{itemize}[leftmargin=*]
\tightlist
    \item \textbf{Secure beamforming and null steering} optimize beamforming weights to maximize legitimate user reception while creating nulls toward eavesdroppers, spatially filtering signals~\cite{thapa2025,RISbasedPLS}.
    
    \item \textbf{Artificial Noise (AN) injection} transmits designed noise orthogonal to the legitimate channel but non-orthogonal to eavesdroppers, degrading their reception~\cite{aman2025,xu2022}.
    
    \item \textbf{Secure waveform design} employs randomized parameters, spread-spectrum, and frequency hopping to make spoofing difficult~\cite{cigno2025,Nguyen}.
    
    \item \textbf{Sensing-assisted security} uses ISAC radar to detect and locate passive eavesdroppers, enabling precise beamforming nulls or artificial noise towards them~\cite{thapa2025,matsumine2025}. 
\end{itemize}
Cost/benefit analysis: The cost involves AN increase in power consumption and a trade-off in communication rate~\cite{PLS_AN}. The benefit is the provision of zero-latency protection against eavesdropping, especially when combined with sensing-assisted adversary localization~\cite{thapa2025,PLS_AN}. Furthermore, leveraging techniques like Constructive Interference (CI) and Destructive Interference (DI) offers an energy-efficient alternative to pure AN injection by steering legitimate signals into decodable regions while actively disrupting unauthorized reception~\cite{aman2025}.

\subsubsection{Pillar 2}
\textbf{Environment-aware security} leverages intelligent technologies, such as Reconfigurable Intelligent Surfaces (RISs), to dynamically shape and control the radio environment in real time~\cite{Naeem2023,thapa2025}. While RIS primarily serves to enhance legitimate communication and suppress eavesdroppers, the trustworthiness of the RIS must be continuously verified to counteract threats such as Malicious/Illegal RIS (IRIS) Injection~\cite{Naeem2023}.
\begin{itemize}[leftmargin=*]
\tightlist
    \item \textbf{RIS-based secure zones} are achieved by optimizing RIS phase shifts for constructive interference for legitimate users and destructive interference for eavesdroppers~\cite{aman2025,RISbasedPLS,PLS_AN}, offering dynamic adaptation to changing threats while balancing communication and sensing~\cite{hafeez2025,meng2024}.
    
    \item \textbf{RIS vulnerability mitigation} addresses the programmable nature of RIS through Cryptographic RIS Control Authentication (e.g., Post-Quantum Cryptography (PQC) like ML-KEM and Falcon)~\cite{hafeez2025} and periodic RIS Integrity Verification (such as GLRT-based scene authentication validated by the Marcum-Q function)~\cite{hafeez2025} to prevent malicious phase shift injections~\cite{aman2025,Naeem2023}. 
    These protections introduce latency but enable dynamic environment optimization without extra transmit power (Passive RIS). 
    
\end{itemize}
Cost/benefit analysis: Reconfiguring RIS introduces additional latency, including cryptographic and settling delays~\cite{hafeez2025}. The benefit is that it dynamically optimizes the radio environment~\cite{RIS_basedISAC}, enhancing legitimate SINR and reducing eavesdropping risk~\cite{matsumine2025}, without needing extra transmit power (Passive RIS)~\cite{RISbasedPLS}.

\subsubsection{Pillar 3}
\textbf{Intelligence-aware security} is key to future networks, offering proactive detection and defense against AI-targeted attacks~\cite{thapa2025,Nguyen}. AI/ML models are crucial for building adaptive defenses, identifying threats missed by rule-based systems~\cite{thapa2025}.
\begin{itemize}[leftmargin=*]
\tightlist
    \item \textbf{Anomaly detection using Deep Learning} detects malicious activities early to enhance security~\cite{aman2025,keskin2025}. 
    Convolutional Neural Networks (CNNs) process data from Range-Doppler Maps (RDM) and CSI to differentiate authorized from unauthorized activities, aiming for real-time detection~\cite{ISAC_types,keskin2025}. 
    
    \item \textbf{Adversarial robustness and certified defenses} improve resilience against evasion attacks, often via adversarial training, retraining models on perturbed~\cite{surve2025}. The objective includes establishing certified robustness, which provides assurance against specified perturbations.
    
    \item \textbf{Federated Learning} enables privacy-preserving, distributed threat detection among agents without sharing raw data~\cite{son2025,bazzi2024}. It reduces computational load~\cite{bazzi2024} and maintains reliable distributed intelligence by filtering outliers, helping resist AI-targeted attacks like evasion and poisoning~\cite{son2025,thapa2025}.
\end{itemize}
Cost/benefit analysis: The primary benefit is enabling proactive, adaptive real-time threat detection~\cite{thapa2025,aman2025}. The cost involves the inherent computational complexity of deploying complex machine learning models, especially when implementing them on resource-constrained devices~\cite{aman2025,bazzi2024}.

\subsubsection{Pillar 4}
\textbf{Architecture-aware security}, based on zero-trust architecture~\cite{ericsson_blog}, enforces continuous verification for all nodes and data flows, replacing the outdated perimeter security model~\cite{Huang_2025}.
\begin{itemize}[leftmargin=*]
\tightlist
    \item \textbf{Zero-trust principles} require continuous authentication (never trust, always verify), granting minimum permissions (least privilege), assuming compromise (assume breach), and verifying explicitly using all available data (identity, location, device health)~\cite{Huang_2025}.
    
    \item \textbf{Continuous authentication} uses metrics based on identity, behavior, and device health to make real-time trust decisions~\cite{Huang_2025}. If a node's score falls below a threshold, the system can enforce Dynamic Revocation or isolation~\cite{keskin2025}.
    
    \item \textbf{Insider threat mitigation} requires securing sensitive ISAC processing, as broadcasting information such as CSI can reveal critical private details~\cite{bazzi2024}. Unsecured aerial links threaten control and sensing channels~\cite{hafeez2025}, necessitating mechanisms such as cryptographic binding of physical-layer features to authentication systems for node verification~\cite{hafeez2025,keskin2025}.
    
\end{itemize}
Cost/benefit analysis: The cost involves latency and computational overhead associated with continuous verification, especially when integrating Post-Quantum Cryptography (PQC) primitives~\cite{keskin2025}. However, this PQC overhead can be manageable relative to overall frame payloads~\cite{hafeez2025}. The benefit is a reduction in insider threats and compromised edge devices through continuous verification~\cite{Huang_2025}, which can also lead to a significant reduction in overall processing costs by using lighter initial authentication checks~\cite{keskin2025}.

\subsection{Cross-layer orchestration and synergy}
The layers coordinate actions bidirectionally, operating within a stringent total orchestration cycle latency of less than $50$ms required for critical infrastructure control loops. For instance, the intelligence layer detects an anomaly and estimates the location, informing the physical layer to compute null steering~\cite{aman2025,wei2022}. The architecture layer then uses trust scores to prevent compromised nodes from transmitting false commands to the physical layer~\cite{5Gamerica_whitepaper}.
The framework emphasizes synergistic coordination, in which the combined effect of multiple layers exceeds the sum of their parts (complementary)~\cite{5Gamerica_whitepaper}.

\textbf{Complementary defenses (synergy):} Intelligence informs the physical layer (e.g., AI detection identifies threat location, physical layer computes null steering towards that direction)~\cite{aman2025,wei2022}. The RIS (environment layer) optimizes signal reception, thereby improving sensing quality at the physical layer, which in turn enhances AI detection accuracy (intelligence layer)~\cite{RIS_basedISAC}. Architecture layer’s zero-trust verdict uses the trust score computation for dynamic revocation, preventing transmission of false commands to the physical/environment layers~\cite{keskin2025,5Gamerica_whitepaper}.

\textbf{Redundant defenses:} While not explicitly categorized as redundant in the sources, the concept of diminishing returns applies when multiple defenses address the exact physical mechanism without coordination (e.g., numerous beamforming strategies that yield only limited marginal security gains). Existing work often treats security layers in isolation, lacking a coordinated approach~\cite{keskin2025}. The effectiveness of the security mechanisms depends on integrating information from different layers rather than relying on single, isolated solutions~\cite{keskin2025}.

\section{Security-performance trade-offs}
Security is not free. It consumes resources (energy, spectrum, time) that could otherwise be used for communication or sensing. ISAC faces the fundamental challenge of jointly optimizing communication rate ($R$), sensing accuracy/distortion ($D_s$), and security (information leakage $I_\textup{leak}$)~\cite{ISAC_survey1}. This joint optimization leads to quantifiable trade-offs characterized by the achievable Rate-Distortion-Leakage region ($R_\textup{RDL}$)~\cite{liu2022}. 

\subsection{Sensing, communication and security trade-offs}
\textbf{Sensing-communication trade-off:} Joint beamforming design must find a Pareto frontier between maximizing communication rate and maximizing sensing accuracy~\cite{ISAC_survey1}. Optimization purely for communication leads to poor sensing due to weak side-lobe illumination of targets (high $D_s$). Conversely, omnidirectional radiation for sensing results in a poor communication rate ($R$)~\cite{PLS_AN}.
    
\textbf{Security-sensing trade-off:} Injecting artificial noise enhances security against eavesdropping~\cite{RISbasedPLS} but degrades sensing performance~\cite{keskin2025}, where performance metrics like the Cramér–Rao Bound (CRB) and capacity–distortion (C–D) are used for theoretical analysis of these trade-offs~\cite{liu2022,cui2025}.

\subsection{Latency-security trade-offs}
Critical ISAC applications require high reliability and low latency~\cite{5Gamerica_whitepaper}, especially for certain specialized use cases~\cite{huawei_wp}. Deploying security mechanisms introduces complexity and computation overhead, including latency, compared to the ideal scenario~\cite{aman2025}.
\begin{itemize}
\tightlist
    \item \textbf{RIS reconfiguration latency:} Deploying security enhancements through RIS necessitates mechanisms to manage latency introduced by cryptographic and settling delays, while accommodating RIS hardware constraints such as switching rate and dwell time, to ensure robust authentication~\cite{hafeez2025}.
    
    \item \textbf{AI inference and training:} AI anomaly detection must be carried out in real time~\cite{wen2025}. However, computational complexity increases when developing resilient models. Distributed training for threat detection utilizes collaborative intelligence by aggregating updates from multiple nodes~\cite{5Gamerica_whitepaper}, which requires robust aggregation rules to manage compromised nodes.
    
    \item \textbf{Zero-trust revocation:} Robust architectural approaches require continuous verification for all nodes and data flows, supporting functions like continuous authentication to defend against insider threats~\cite{ericsson_wp}.
\end{itemize}

\subsection{Context-dependent optimization}
Optimization strategies must be tailored based on application priorities. The complex interplay requires jointly optimizing communication rate, sensing accuracy, and security, often characterized by the achievable Rate-Distortion-Leakage region or by solving multi-objective optimization problems~\cite{meng2024,liu2022}.

\textbf{Autonomous vehicle on highway (priority: low latency > sensing accuracy > security):} Configuration employs strategies that favor real-time control. Autonomous driving applications require ultra-reliable, low-latency communications (URLLC), often targeting millisecond latency. To achieve the highest communication rate ($R$), aggressive beamforming is crucial~\cite{bazzi2024}. However, security mechanisms, such as those that enforce continuous verification (Zero-Trust), can introduce latency. Additionally, integrating artificial noise to enhance security against eavesdropping may degrade sensing performance~\cite{keskin2025}.
    
\textbf{Critical power grid SCADA (priority: security > sensing accuracy > communication rate):} Configuration requires mechanisms that place a high priority on safeguarding the system. Critical infrastructure applications prioritize resilience and integrity~\cite{5Gamerica_whitepaper}. This requires continuous defense measures, including enforcing continuous verification (Zero-Trust Architecture) of all nodes~\cite{thapa2025}. Mitigating insider threats necessitates securing the processing of sensitive information, such as Channel State Information (CSI)~\cite{bazzi2024}. The relative stability of large-scale systems means they do not always require the extreme ultra-low latency required for applications such as autonomous driving~\cite{wang2025}.
    
\textbf{Healthcare vital signs monitoring (priority: privacy > security > accuracy):} Configuration relies on measures that minimize the exposure of sensitive health data, as privacy is a key requirement for such systems~\cite{Nguyen}. This is achieved by implementing on-device processing to reduce the communication overhead of data transmission~\cite{kajol2025}. Privacy preservation is enhanced by using Federated Learning (FL), which enables distributed, privacy-preserving model training without exchanging raw data. FL is often used in conjunction with Differential Privacy (DP) schemes, which introduce randomness into data prior to transmission to prevent inference of information~\cite{Naeem2023}. However, in critical applications like remote robotic surgery, the priority must shift to ultra-reliable, low-latency communication (URLLC) to ensure safety and reliability.

\section{Critical infrastructure sector-specific requirements}
The security requirements for ISAC systems vary significantly across critical infrastructure sectors, reflecting diverse potential physical consequences of an attack~\cite{thapa2025}. Cyber-physical systems, such as smart grids, face sophisticated threats in which consequences can be complex and difficult to mitigate, potentially leading to catastrophic physical harm~\cite{Huang_2025}.

\subsection{Smart grids}

\textbf{Threat model:} Deceptive jamming poses a risk to SCADA sensing systems by aiming to mislead the victim radar receiver by generating phantom targets~\cite{target_spoofing}, creating phantom dangers~\cite{keskin2025}, or exploiting the inherent civilian susceptibility of sensing mechanisms to inaccuracies~\cite{Privacysecurity1}. Furthermore, adversaries may utilize CSI manipulation attacks by injecting subtle interference to alter CSI~\cite{thapa2025}. Such manipulation could mislead the perception systems, potentially biasing voltage measurements (monitored by PMUs)~\cite{surveysecuritysmart}.
    
\textbf{Impact severity:}  The impact is categorized as critical, as disruptions can lead to grid-wide cascading blackouts~\cite{surveysecuritysmart}.
    
\textbf{Required defenses:} Defenses must be multi-layered to create a resilient ecosystem~\cite{thapa2025}. This necessitates implementing continuous zero-trust verification protocols, as architecture-aware security enforces continuous verification of all nodes and data flows, countering obsolete perimeter security models~\cite{thapa2025}. Robust defense strategies also involve injecting artificial noise, which intentionally degrades eavesdroppers’ channel quality to suppress wiretapping without significantly affecting legitimate users~\cite{xu2022}. Furthermore, solutions must incorporate Byzantine robustness to maintain system functionality even if specific cooperating nodes are compromised by filtering outlier updates~\cite{aman2025}. The required system latency for critical infrastructure control loops is considered stringent~\cite{bazzi2024}, and overall ISAC applications face an approximate end-to-end latency tolerance of the order of milliseconds~\cite{ISAC_survey1}.
\vspace{-0.2cm}

\subsection{Intelligent transportation systems}
Intelligent transportation systems and autonomous driving are core application scenarios for ISAC~\cite{keskin2025,matsumine2025}.

\textbf{Threat model:} Target spoofing attacks aim to mislead a victim radar receiver by injecting false data to create phantom targets or phantom obstacles~\cite{keskin2025,thapa2025}. Deceptive jamming can render real obstacles invisible by manipulating the perceived attributes of a real target~\cite{keskin2025,thapa2025}.
    
\textbf{Impact severity:} The consequences of insecure ISAC in transportation are critical~\cite{keskin2025}. Cyber-physical attacks that corrupt sensing data can lead to misjudgments and safety incidents in autonomous driving~\cite{geng2024}, potentially resulting in unsafe actuation~\cite{keskin2025} and serious accidents~\cite{Privacysecurity1}, such as multi-vehicle collisions or loss of life~\cite{Huang_2025}.
    
\textbf{Required defenses:} Transportation systems demand ultra-reliable low-latency communications (URLLC)~\cite{bazzi2024}. Defenses should include strong sensor fault tolerance and redundancy, exploiting ISAC's ability to provide infrastructure-based sensing to cross-verify onboard sensor data. This external input can help resolve conflicting observations, such as alerting a vehicle if its own radar has been spoofed~\cite{keskin2025}.
\vspace{-0.2cm}

\subsection{Smart healthcare}
\textbf{Threat model:} The convergence of ISAC and healthcare applications makes systems vulnerable to the manipulation of sensitive physiological states, such as respiration and heart rate monitoring~\cite{geng2024}. Spoofing sensing data, or introducing false data to compromise sensing accuracy, can lead to misjudgments that cause safety incidents~\cite{keskin2025}. Furthermore, wireless signals, including mmWave reflections, can be exploited by adversaries who maliciously estimate sensitive information about physiological states through passive eavesdropping on reflected signals~\cite{geng2024}.
    
\textbf{Impact severity:} Healthcare applications, such as remote robotic surgery, are considered mission-critical~\cite{bazzi2024}. Security breaches in critical infrastructure, including applications in smart cities and healthcare, pose considerable risks to public safety~\cite{Huang_2025}.
    
\textbf{Required defenses:} The highest priority must be placed on privacy and security, as they are fundamental for the medical sector~\cite{cigno2025}. Therefore, sensing needs to be performed in a manner that prioritizes privacy assurance~\cite{ericsson_wp}. This requires strict adherence to regulations such as the GDPR, which enforces rules on data processing~\cite{PLS_AN}. Key compliance requirements include the principle of data minimization, which mandates that personal data collection and processing be limited to what is strictly necessary, and purpose limitation, ensuring that collected personal data is not processed incompatibly with initially specified, explicit, and legitimate purposes~\cite{Privacysecurity2}. Technical solutions include using Federated Learning (FL) to enable distributed, privacy-preserving model training, often complemented by Differential Privacy (DP) schemes that add randomness to data to prevent inference of information~\cite{geng2024}. Furthermore, securing the integrity of sensitive information requires using mechanisms such as cryptographic binding of physical-layer features with authentication systems~\cite{Guo2024}.

\subsection{Unmanned aerial vehicles, industrial systems and smart cities}

\textbf{Unmanned Aerial Vehicles (UAVs):} UAV systems face security risks such as spoofing, hijacking of control channels, and attacks on trajectory prediction~\cite{hafeez2025}. As UAV corridors require highly efficient communication and decentralized coordination, aerial links are susceptible to eavesdropping and interference, endangering control and sensing channels. Ensuring secure UAV systems requires ultra-reliable low-latency communications (URLLC), though compliance with required response times remains challenging due to cryptographic and settling delays~\cite{wen2025,hafeez2025}.
    
\textbf{Industrial systems:} The focus is on implementing high safety integrity where cyber-physical attacks could result in catastrophic physical harm~\cite{thapa2025}. ISAC applications in sectors like intelligent manufacturing require ultra-reliable, low-latency performance~\cite{wen2025}. Applications such as module installation and placement in factories and warehouses require extremely high accuracy, sometimes approaching centimeter-level precision~\cite{huawei_wp}. Given that Operational Technology (OT) prioritizes safety and reliability, robust security models are critical for industrial control systems~\cite{Huang_2025}.
    
\textbf{Smart cities:} These environments face severe risks of sensing eavesdropping, where unauthorized receivers exploit sensing signals to infer precise locations, movements, or activities of individuals~\cite{thapa2025}. The potential for misuse of ISAC technology to enable a mass surveillance scenario~\cite{thapa2025}. Priority is placed on high-privacy protection. Technical mechanisms for dual-use prevention must be implemented, adhering to security principles such as data minimization and purpose limitation~\cite{Privacysecurity2}. Solutions often involve ensuring user consent and creating immutable audit logging by ensuring that a disclosure log records recipient identities, data descriptions, and timestamps~\cite{Privacysecurity2}.

\section{Research gaps and emerging challenges}

\subsection{Critical research gaps}

\textbf{Information-theoretic foundations:} The fundamental limits on the joint optimization of communication rate ($R$), sensing accuracy ($D_s$), and security ($I_\textup{leak}$) remain theoretically underdeveloped~\cite{cui2025,liu2022,ISAC_survey1}. Open problems include characterizing the Rate-Distortion-Leakage Region and analyzing security under active threat scenarios (e.g., an eavesdropper that can also jam).\\
\textbf{Cross-layer orchestration protocols:} While the framework specifies layers and stringent timing requirements, the development of formal protocols for real-time coordination across distributed nodes is recognized as a critical challenge, as cross-layer optimization for ISAC networks is still in its infancy~\cite{ISAC_survey1}. Current security models often employ simplistic approaches that fail to adequately account for the dynamic, multi-layer, multi-path, and multi-agent characteristics prevalent in real-world Cyber-Physical Systems (CPS)~\cite{Huang_2025}.\\
\textbf{Certified adversarial robustness for RF signals:} Robustness is well-studied for images, but nascent for radio frequency signals. Open problems include defining the physically realizable threat model for RF adversarial examples (accounting for hardware constraints) and computing the certified robustness radius for RF anomaly detectors~\cite{surve2025}. Defining robust privacy metrics for ISAC systems is challenging because traditional Differential Privacy (DP) can be arbitrarily weak when sensitive sensing data exhibit high correlation, necessitating the use of advanced concepts such as Pointwise Maximal Leakage (PML) to achieve certified robustness guarantees~\cite{sara}.\\
\textbf{Privacy-by-design architecture:} Integration of Privacy-Enhancing Technologies (PETs) with ISAC security is incomplete. Gaps include developing differentially private sensing techniques and exploring the feasibility of Homomorphic Encryption for ISAC (processing encrypted sensing data in real time)~\cite{Privacysecurity2}\cite{PLS_AN}.

\subsection{Emerging challenges}
\textbf{Quantum-resilient ISAC:} The looming threat of quantum computing breaking classical asymmetric cryptography necessitates research into quantum-resilient ISAC. Key challenges include designing efficient Quantum-Safe Key Establishment protocols for ISAC, accounting for bandwidth and latency constraints, and developing Quantum-Resilient Sensing Signatures~\cite{hafeez2025}. This includes addressing the threat posed by Harvest-Now Decrypt-Later (HNDL) attacks using Post-Quantum Cryptography (PQC).\\
\textbf{Advanced AI attacks:} Future threats involve adaptive attacks that learn and exploit system defenses. Required actions include migrating to systems with certified robustness, developing continuous learning systems that adapt faster than attackers, and increasing AI interpretability (XAI) for anomaly analysis~\cite{son2025}.

\begin{table*} [t]
    \centering
    \vspace{-0.2cm}
    \caption{Key related works.}
    \scriptsize
    \label{tab:survey}
    \begin{tabular}{|p{1.5cm}|p{2cm}|p{3.2cm}|p{1.4cm}|p{3.5cm}|}
    \hline
    \multicolumn{1}{|c|}{\textbf{Work}} &
    \multicolumn{1}{|c|}{\textbf{Type} }&
    \multicolumn{1}{|c|}{\textbf{Scope} } &
    \multicolumn{1}{|c|}{\textbf{Depth} }&
    \multicolumn{1}{|c|}{\textbf{Novelty}} 
       \\ \hline
          Thapa et al.~\cite{thapa2025} & Vision paper & Perception-centric security, defense-in-depth & Conceptual & Identified key functional transformations.
          \\ \hline
          Matsumine et al.~\cite{PLS1} & Review & Physical layer security & High & PLS mechanisms. 
         \\ \hline
          Li et al.~\cite{RISbasedPLS} & Review & RIS-based security & Moderate & RIS-specific solutions. 
         \\ \hline
          Su et al.~\cite{PLS_AN} & Review & Multi-layered defense-in-depth framework and privacy & High & Systematic security framework addressing the integrity of perception.
         \\ \hline
          Qu et al.~\cite{Privacysecurity1} &  Review & Privacy in ubiquitous sensing & Moderate & Privacy-centric view. 
         \\ \hline
         Lu et al. ~\cite{ISAC_survey1} & Comprehensive Review & Foundations, physical-layer system design, networking aspects & High & Framework organizing ten open challenges across the theoretical foundation, physical-layer system design, networking, and application layers.
         \\ \hline
         Martins et al. & Review & Security and privacy overview & Moderate & Threat taxonomy (basic).
         \\ \hline
          \textbf{This work} & Comprehensive review & Multi-dimensional analysis: threats, vulnerabilities, defenses, trade-offs, requirements and gaps &  High & Threat taxonomic framework across multiple dimensions. 
         \\ \hline    
      
    \end{tabular} 
    \vspace{-0.2cm}
\end{table*}
\vspace{-0.3cm}
\section{Related works}
Existing security research relevant to ISAC falls into several categories, addressing the foundational theory, practical mechanisms, and specific threat models:\\
\textbf{(1) Foundational concepts:} Works cover ISAC fundamentals (\cite{liu2022}), joint radar-communication design (\cite{cui2025}), and physical layer security baseline (\cite{matsumine2025}).\\
\textbf{(2) PLS and defense mechanisms:} Many papers focus on implementing PLS using methods such as artificial noise injection, cooperative jamming, and secure beamforming~\cite{thapa2025,matsumine2025,aman2025}. Solutions aim to maximize the Secrecy Rate (SR) for robust transmission against targets acting as eavesdroppers.\\
\textbf{(3) RIS-assisted security:} Numerous studies investigate the use of Reconfigurable Intelligent Surfaces (RIS) to enhance security by reflecting signals destructively toward eavesdroppers~\cite{RISbasedPLS,aman2025}, including optimizing SR and sensing Signal-to-Noise Ratio (SNR). \\
\textbf{(4) Threat and vulnerability analysis:} Research identifies vulnerabilities from a perception-centric viewpoint, focusing on security and privacy challenges in ubiquitous sensing~\cite{cigno2025,geng2024}. Specific attack models include the dual threat of spoofing and jamming in OFDM/WLAN sensing~\cite{target_spoofing} and adversarial attacks on machine learning models~\cite{geng2024}.

\section{Conclusion}
The integration of Integrated Sensing and Communication (ISAC) is transforming perceptive networks and defining the direction of sixth-generation (6G) technologies. This paper introduces a detailed framework for classifying and examining the full spectrum of security challenges in ISAC. The framework systematically identifies potential threats and layered vulnerabilities across design, physical, computational, and architectural levels. It also highlights a defense-in-depth strategy based on four interconnected pillars: physical, environmental, intelligence, and architectural security. Achieving trustworthy ISAC deployment requires carefully balancing security and performance trade-offs while addressing emerging challenges, including the certified adversarial robustness of radio frequency signals, quantum resilience, and privacy protection. By providing a comprehensive analysis and identifying key research gaps, this work establishes a foundational reference for securing next-generation perceptive networks.

%
%
%
{\tiny
\let\oldthebibliography\thebibliography
\let\endoldthebibliography\endthebibliography
\renewenvironment{thebibliography}[1]{
  \oldthebibliography{#1}
  \setlength{\itemsep}{0pt}
  \setlength{\parsep}{0pt}
}{\endoldthebibliography}
\bibliographystyle{splncs04}
\bibliography{bibliography}}
\end{document}